\documentclass[prx, preprint, superscriptaddress]{revtex4-2}
\usepackage{amsmath}
\usepackage{amssymb}
\usepackage{graphicx}
\usepackage[caption=false, position=top, singlelinecheck=off, justification=raggedright]{subfig}
\usepackage{multirow}
\usepackage{color}
\usepackage{pst-node}
\usepackage[unicode]{hyperref}
\hypersetup{
	unicode=true,         	
	colorlinks=true,      	
	linkcolor=blue,		   	
	citecolor=blue,       	
	urlcolor=blue		   	
}

\begin{document}
	
\title{Parametric control of Meissner screening in light-driven superconductors}

\author{Guido Homann}
\affiliation{Zentrum f\"ur Optische Quantentechnologien and Institut f\"ur Laserphysik, 
	Universit\"at Hamburg, 22761 Hamburg, Germany}

\author{Jayson G. Cosme}
\affiliation{National Institute of Physics, University of the Philippines, Diliman, Quezon City 1101, Philippines}

\author{Ludwig Mathey}
\affiliation{Zentrum f\"ur Optische Quantentechnologien and Institut f\"ur Laserphysik, 
	Universit\"at Hamburg, 22761 Hamburg, Germany}
\affiliation{The Hamburg Centre for Ultrafast Imaging, Luruper Chaussee 149, 22761 Hamburg, Germany}

\date{\today}
\begin{abstract}
We investigate the Meissner effect in a parametrically driven superconductor using a semiclassical $U(1)$ lattice gauge theory. Specifically, we periodically drive the $z$-axis tunneling, which leads to an enhancement of the imaginary part of the $z$-axis conductivity at low frequencies if the driving frequency is blue-detuned from the plasma frequency. This has been proposed as a possible mechanism for light-enhanced interlayer transport in YBa$_2$C$_3$O$_{7-\delta}$ (YBCO). In contrast to this enhancement of the conductivity, we find that the screening of magnetic fields is less effective than in equilibrium for blue-detuned driving, while it displays a tendency to be enhanced for red-detuned driving.
\end{abstract}
\maketitle

\newpage
\section{Introduction}
\label{sec:intro}
Optical driving of solids opens up the possibility to induce superconducting-like features in their response to electric fields. This was first achieved in several cuprates by the excitation of specific phonon modes \cite{Fausti2011, Hu2014} or near-infrared excitation \cite{Nicoletti2014, Cremin2019}. Later, signatures of a superconducting state were induced in fullerides and organic salts by exciting molecular vibrations \cite{MItrano2016, Budden2021, Buzzi2020}.
In all these experiments, the imaginary part $\sigma_2 (\omega)$ of the optical conductivity exhibited a $1/\omega$ divergence at low frequencies following optical excitation at temperatures above the equilibrium critical temperature $T_c$. In the case of YBCO, an enhancement of the low-frequency conductivity $\sigma_2 (\omega)$ along the $c$ axis was also observed below $T_c$ \cite{Hu2014, Kaiser2014}. Several mechanisms have been proposed to explain the enhancement of interlayer transport, including nonlinear lattice dynamics \cite{Mankowsky2014}, parametric driving \cite{Denny2015, Hoeppner2015, Okamoto2016, Michael2020}, and suppression of competing orders \cite{Raines2015, Patel2016}.
While the transient optical response of the light-driven cuprates and organic materials is consistent with enhanced or induced superconducting states, their response to magnetic fields has remained largely unexplored. That is due to the limited lifetimes of the excited states, which make experimental measurements of the magnetic response challenging \cite{Paone2021}.
Therefore, it is an open question whether the experimental observations of the light-induced transport properties indeed correspond to light-enhanced or light-induced superconductivity in the sense of an enhanced Meissner effect \cite{Chiriaco2018, Bittner2019, Paeckel2020, Dai2021a, Dai2021b}.

In this paper, we theoretically study the Meissner effect in a parametrically driven superconductor. We consider a specific mechanism of parametric driving, where the Cooper pair tunneling along the $z$ axis is periodically modulated in time \cite{Okamoto2016, Okamoto2017}. This type of driving enhances the imaginary part of the optical conductivity along the $z$ axis at low frequencies. Based on analytical and numerical calculations, we find that the screening of DC magnetic fields is less effective than in equilibrium for slightly blue-detuned driving. This is due to the generation of electromagnetic waves by the parametric driving.
For red-detuned parametric driving, there is no transmission of electromagnetic waves into the bulk and the Meissner screening is enhanced on a length scale that depends on the driving strength and the driving frequency at the order of our analytical investigation. The enhancement of the Meissner screening is particularly effective when the driving frequency is close to the plasma frequency. Notably, the imaginary part of the optical conductivity is reduced in this regime of driving frequencies.

This paper is organized as follows. After introducing our semiclassical method in Section~\ref{sec:Method}, we discuss the optical conductivity of a parametrically driven superconductor in Section~\ref{sec:Conductivity}. In Section~\ref{sec:Meissner}, we first investigate the Meissner effect in a parametrically driven superconductor from an analytical perspective. Furthermore, we present numerical results for parametrically driven superconductors with isotropic and anisotropic lattice parameters. We conclude this work in Section~\ref{sec:Conclusion}.

\begin{figure}[!b]
	\centering
	\includegraphics[scale=1]{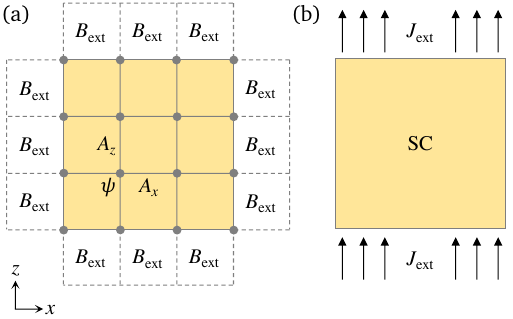}
	\caption{Magnetic and optical probing of a superconductor (SC). (a) Setup of the lattice gauge simulation with a fixed magnetic field at the surface. The order parameter is located on the lattice sites, while the vector potential is defined on the bonds. (b) The $z$-axis optical conductivity is measured by applying a spatially uniform probe current and setting $B_{\mathrm{ext}}=0$.}
	\label{fig:fig1}
\end{figure}

\section{Method}
\label{sec:Method}
Here, we give an overview of the semiclassical $U(1)$ lattice gauge theory that we utilize to simulate the dynamics of a parametrically driven superconductor \cite{Homann2020, Homann2021, Homann2022}.
The static part of the Lagrangian is the Ginzburg-Landau free energy \cite{Ginzburg1950} on a three-dimensional lattice. As depicted in Fig.~\ref{fig:fig1}(a), the superconducting order parameter $\psi_{\mathbf{r}}(t)$ is located on the sites of a cubic lattice with lattice constant $d$, where $\mathbf{r}= (x,y,z)$ is the lattice site. The components of the electromagnetic vector potential $A_{j,\mathbf{r}}(t)$ are defined on the lattice bonds, which connect each site $\mathbf{r}$ with its nearest neighbor in the $j \in \{x,y,z\}$ direction. We employ the temporal gauge such that the electric field components are calculated according to $E_{j,\mathbf{r}} = -\partial_{t} A_{j,\mathbf{r}}$. The magnetic field components $B_{j,\mathbf{r}}= \epsilon_{jkl} (A_{l,\mathbf{r'}(k)}-A_{l,\mathbf{r}})/d$ are found on the lattice plaquettes, with $\mathbf{r'}(k)$ as the neighboring site of $\mathbf{r}$ in the $k$~direction.

The Lagrangian of the lattice gauge model is
\begin{align} \label{eq:Lagrangian}
\begin{split}
	\mathcal{L} ={}& \sum_{\mathbf{r}} K \hbar^2 | \partial_{t} \psi_{\mathbf{r}} |^2 + \mu | \psi_{\mathbf{r}} |^2 - \frac{g}{2} | \psi_{\mathbf{r}} |^4 - \sum_{j,\mathbf{r}} t_j |\psi_{\mathbf{r'}(j)} - \psi_{\mathbf{r}} e^{i a_{j,\mathbf{r}}}|^2 \\
	&+ \sum_{j,\mathbf{r}} \frac{\epsilon_0}{2} E_{j,\mathbf{r}}^2 - \frac{\hbar^2}{4 \mu_0 e^2 d^4} \left[1 - \cos\left(\frac{2ed^2}{\hbar} B_{j,\mathbf{r}} \right) \right] ,
\end{split}
\end{align}
where $\mu$ and $g$ are the Ginzburg-Landau coefficients and the coefficient $K$ describes the magnitude of the dynamical term \cite{Pekker2015,Shimano2020}. The dynamical term is of the form $|\partial_t \psi_{\mathbf{r}}|^2$, which supports the particle-hole symmetry of the Lagrangian, i.e., $\mathcal{L}$ is invariant under $\psi_{\mathbf{r}} \rightarrow \psi_{\mathbf{r}}^*$ and $e \rightarrow -e$.
The coupling of the unitless vector potential $a_{j,\mathbf{r}}= -2ed A_{j,\mathbf{r}}/\hbar$ to the phase of the order parameter ensures the local gauge-invariance of the Lagrangian. Note that the charge of a Cooper pair is $-2e$. The tunneling coefficients $t_x= t_y= t_{xy}$ and $t_z$ determine the plasma frequencies of the superconductor,
\begin{equation} \label{eq:plasmaFreq}
	\omega_j= \sqrt{\frac{8t_j n_0 e^2 d^2}{\epsilon_0 \hbar^2}} ,
\end{equation}
with the equilibrium Cooper pair density $n_0= \mu/g$. The superconductor is isotropic for $t_{xy}= t_z$ and anisotropic for $t_{xy} \neq t_z$.

We derive the Euler-Lagrange equations from Eq.~\eqref{eq:Lagrangian} and include damping terms,
\begin{align}
	\partial_t^2 \psi_{\mathbf{r}} &= \frac{1}{K \hbar^2} \frac{\partial \mathcal{L}}{\partial \psi_{\mathbf{r}}^*} - \gamma_{\mathrm{sc}} \partial_t \psi_{\mathbf{r}} , \\
	\partial_t^2 A_{j, \mathbf{r}} &= \frac{1}{\epsilon_0} \frac{\partial \mathcal{L}}{\partial A_{j, \mathbf{r}}} - \gamma_{\mathrm{el},j} \partial_t A_{j, \mathbf{r}} ,
\end{align}
where $\gamma_{\mathrm{sc}}$ and $\gamma_{\mathrm{el},j}$ are phenomenological damping coefficients of the order parameter and the electric field, respectively. We note that these equations are the zero-temperature limit of the Langevin equations used in Ref.~\cite{Homann2022}.

We numerically solve the equations of motion employing periodic boundary conditions along the $y$ axis. With this boundary condition, we take the superconducting sample to be spatially homogeneous along the $y$ axis, rather than having open boundary conditions. We assume open boundary conditions in the $x$ and $z$ direction and impose a spatially uniform magnetic field $\mathbf{B}= B_{\mathrm{ext}} \mathbf{\hat{y}}$ at the surfaces in $x$ and $z$ direction. For this purpose, we add one numerical layer outside the sample; see Fig.~\ref{fig:fig1}(a). On the external plaquettes, we fix the magnetic field to zero, ramp it up to a non-zero constant or specify a temporally oscillating value in the following. Thus, we model different physical scenarios. To simulate the vacuum, we set the order parameter and the tunneling coefficients to zero outside the sample. Inside the sample, we initialize the order parameter and the vector potential in the ground state, where $\psi_{\mathbf{r}} \equiv \sqrt{\mu/g}$ and $\mathbf{A}_{\mathrm{r}} \equiv 0$, and integrate the differential equations using Heun's method with a step size of $\Delta t= 2.5~\mathrm{ns}$.

\section{Conductivity of parametrically driven superconductors}
\label{sec:Conductivity}
We measure the optical conductivity by adding a weak probe current $J_{\mathrm{ext}}(t)= J_0 \cos(\omega_{\mathrm{pr}} t)$ to the equations of motion for the $z$ component of the electric field,
\begin{equation}
	\partial_t^2 A_{z, \mathbf{r}} = \frac{1}{\epsilon_0} \frac{\partial \mathcal{L}}{\partial A_{z, \mathbf{r}}} - \gamma_{\mathrm{el},z} \partial_t A_{z, \mathbf{r}} - \frac{J_{\mathrm{ext}}}{\epsilon_0} ,
\end{equation}
as depicted in Fig.~\ref{fig:fig1}(b). For this measurement, we fix the surface magnetic field to $B_{\mathrm{ext}}=0$, neglecting radiation from the sample due to the probe current. Thus, the dynamics is spatially homogeneous along the $x$ axis and independent of the sample width. The optical conductivity is $\sigma (\omega_{\mathrm{pr}}) = J_{\mathrm{ext}} (\omega_{\mathrm{pr}}) / E (\omega_{\mathrm{pr}})$, where $E (\omega_{\mathrm{pr}})$ is the Fourier transform of the spatial average of the electric field in the steady state.
Additionally, we drive the sample by periodically modulating the tunneling coefficients of all $z$-axis junctions,
\begin{equation} \label{eq:driving}
	t_z \rightarrow t_z \left[ 1+ M \cos(\omega_{\mathrm{dr}}t) \right] .
\end{equation}
Experimentally, this could be achieved by resonantly exciting an infrared-active phonon mode; see Refs.~\cite{Hu2014, Liu2020, Okamoto2016, Okamoto2017}.

\begin{figure}[!t]
	\centering
	\includegraphics[scale=1]{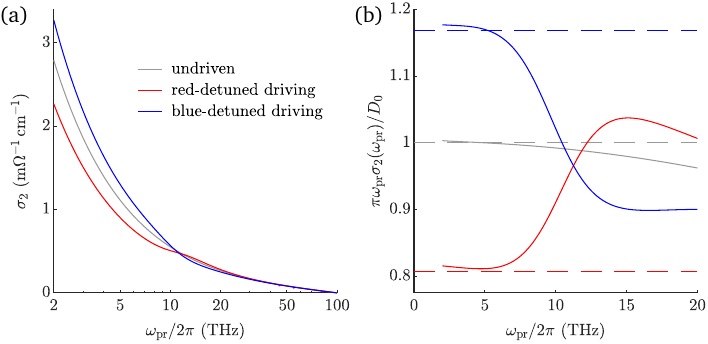}
	\caption{Optical conductivity of a parametrically driven superconductor. (a) Imaginary part $\sigma_2$ of the optical conductivity. (b) $\sigma_2$ multiplied by the probe frequency. The dashed lines indicate the analytical prediction for the superconducting weight, which corresponds to the zero-frequency limit of $\pi \omega_{\mathrm{pr}} \sigma_2(\omega_{\mathrm{pr}})$. The $z$-axis tunneling coefficient is driven with the frequency $\omega_{\mathrm{dr}}= 0.9 \omega_{\mathrm{pl}}$ in the red-detuned case and with the frequency $\omega_{\mathrm{dr}}= 1.1 \omega_{\mathrm{pl}}$ in the blue-detuned case. The driving strength is $M=0.3$ in both cases. The plasma frequency is $\omega_{\mathrm{pl}}/2\pi= 100~\mathrm{THz}$ and the sample height is $100~\mathrm{\mu m}$; see supplementary material for the full parameter set (isotropic sample).}
	\label{fig:fig2}
\end{figure}

The effect of this parametric driving on the imaginary part $\sigma_2(\omega_{\mathrm{pr}})$ of the $z$-axis conductivity is displayed in Fig.~\ref{fig:fig2}(a). While $\sigma_2$ is reduced for $\omega_{\mathrm{dr}} < \omega_{\mathrm{pl}}$ at probe frequencies $\omega_{\mathrm{pr}} \lesssim |\omega_{\mathrm{pl}}- \omega_{\mathrm{dr}}|$, it is enhanced for $\omega_{\mathrm{dr}} > \omega_{\mathrm{pl}}$.
Figure~\ref{fig:fig2}(b) reveals that $\sigma_2$ approaches a $1/\omega_{\mathrm{pr}}$ behavior at low probe frequencies, regardless of whether the superconductor is driven or not. These results are consistent with the findings of Refs.~\cite{Okamoto2016, Okamoto2017}. To quantify the superconducting character of the optical response, we use the superconducting weight
\begin{equation}
	D= \pi \left[ \omega_{\mathrm{pr}} \sigma_2(\omega_{\mathrm{pr}}) \right]_{\omega_{\mathrm{pr}} \rightarrow 0} ,
\end{equation}
following the definition given in Ref.~\cite{Resta2018}.
For an infinitely large sample, the analytical expression for the superconducting weight is $D_0= \pi \epsilon_0 \omega_{\mathrm{pl}}^2$. An analytical prediction for the parametrically driven case was derived in Ref.~\cite{Okamoto2016},
\begin{equation}
	D= D_0 \left(1 - \frac{M^2 \omega_{\mathrm{pl}}^2 (\omega_{\mathrm{pl}}^2 - \omega_{\mathrm{dr}}^2)}{2(\omega_{\mathrm{pl}}^2 - \omega_{\mathrm{dr}}^2)^2 + 2\gamma_{\mathrm{el},z}^2 \omega_{\mathrm{dr}}^2} \right) .
\end{equation}
As one can see in Fig.~\ref{fig:fig2}(b), our numerical results for $M=0.3$ are in good agreement with this prediction. In the blue-detuned case of $\omega_{\mathrm{dr}}= 1.1 \omega_{\mathrm{pl}}$, the superconducting weight is enhanced by approximately $17\%$. In the red-detuned case of $\omega_{\mathrm{dr}}= 0.9 \omega_{\mathrm{pl}}$, the superconducting weight is reduced by approximately $19\%$. We note that the numerically obtained conductivity $\sigma_2$ does not strictly follow the predicted $1/\omega_{\mathrm{pr}}$ behavior due to the insufficiently small probe frequencies and the finite sample size. The effect of parametric driving on the real part of the optical conductivity and the redistribution of spectral weight are discussed in the supplementary material.

\section{Meissner screening in parametrically driven superconductors}
\label{sec:Meissner}

\subsection{Analytical results}
\label{sec:analytical}
The magnetic response of the parametrically driven superconductor can be understood from an analytical perspective. We suppose that the bottom left corner of the superconductor has the coordinates $x=0$ and $z=0$. First, we derive an approximate equation for the dynamics of the magnetic field close to the center of the left surface, i.e., for $x \ll L_x$ and $z=L_z/2$, where $L_x$ and $L_z$ are the edge lengths of the superconductor in $x$ and $z$ direction, respectively.
According to our semiclassical $U(1)$ gauge theory, the supercurrent density along the $z$~axis is given by
\begin{equation}
	J_{z, \mathbf{r}}^{\mathrm{(sup)}} = \frac{2t_z e d}{i \hbar} \left( \psi_{\mathbf{r'}(z)}^* \psi_{\mathbf{r}} e^{ia_{z, \mathbf{r}}} - c.c. \right) \left[1+ M \cos(\omega_{\mathrm{dr}} t) \right] .
\end{equation}
Neglecting fluctuations of the superconducting order parameter, this can be simplified to
\begin{equation}
	J_{z, \mathbf{r}}^{\mathrm{(sup)}} \approx \frac{4t_z n_0 e d}{\hbar} \sin(a_{z, \mathbf{r}}) \left[1+ M \cos(\omega_{\mathrm{dr}} t) \right] .
\end{equation}
In the previous step, we fixed the gauge such that $\arg (\psi_{\mathbf{r}}) \equiv 0$, which complies with the temporal gauge in a charge neutral system.
For weak fields, we linearize the above expression and rewrite it using the expression for the plasma frequency $\omega_{\mathrm{pl}}$ from Eq.~\eqref{eq:plasmaFreq},
\begin{equation}
	J_{z, \mathbf{r}}^{\mathrm{(sup)}} \approx - \epsilon_0 \omega_{\mathrm{pl}}^2 A_{z, \mathbf{r}} \left[1+ M \cos(\omega_{\mathrm{dr}} t) \right] .
\end{equation}
In the following, we treat $\mathbf{A}$ and $\mathbf{J}$ as continuous fields and drop the subscript $\mathbf{r}$. As $A_x \sim J_x = 0$ for $z=L_z/2$, the magnetic field is given by $B_y= - \partial_x A_z$. Neglecting the current contribution from the damping term, we obtain
\begin{equation} \label{eq:rotJ}
	\mathbf{\nabla} \times \mathbf{J} \approx - \epsilon_0 \omega_{\mathrm{pl}}^2 \left[1+ M \cos(\omega_{\mathrm{dr}} t) \right] B_y \mathbf{\hat{y}}
\end{equation}
for the curl of the free current density. On the other hand, Maxwell's equations imply
\begin{equation} \label{eq:combiMaxwell}
	\frac{1}{\mu_0} \left( \frac{1}{c^2} \partial_t^2 - \mathbf{\nabla}^2 \right) \mathbf{B} = \mathbf{\nabla} \times \mathbf{J} .
\end{equation}
Combining Eqs.~\eqref{eq:rotJ} and \eqref{eq:combiMaxwell} yields the minimal model
\begin{equation} \label{eq:minimalModel}
	\partial_t^2 B_y + \omega_{\mathrm{pl}}^2 \left[ 1+ M \cos(\omega_{\mathrm{dr}} t) \right] B_y \approx c^2 \partial_x^2 B_y ,
\end{equation}
as $\partial_z^2 B_y \ll \partial_x^2 B_y$ for $x \ll L_x$ and $z=L_z/2$.
For $M \ll 1$, we use the ansatz
\begin{equation}
	B_y(x,t)= B_0(x) + B_1(x) \cos(\omega_{\mathrm{dr}} t) + B_2(x) \sin(\omega_{\mathrm{dr}} t) .
\end{equation}
This leads to
\begin{align}
	\partial_x^2 B_0 &= \frac{1}{\lambda^2} \left( B_0 + \frac{M}{2} B_1 \right) , \label{eq:B0} \\
	\partial_x^2 B_1 &= \frac{1}{\lambda^2} \left( 1- \frac{\omega_{\mathrm{dr}}^2}{\omega_{\mathrm{pl}}^2} \right) B_1 + \frac{M}{\lambda^2} B_0 , \label{eq:B1} \\
	\partial_x^2 B_2 &= \frac{1}{\lambda^2} \left( 1- \frac{\omega_{\mathrm{dr}}^2}{\omega_{\mathrm{pl}}^2} \right) B_2 , \label{eq:B2}
\end{align}
where $\lambda= c/\omega_{\mathrm{pl}}$ is the London penetration depth of the undriven superconductor. To determine $B_0(x)$ and $B_1(x)$, we use the ansatz
\begin{align}
	B_0 &= b_0 e^{-x/\ell} , \\
	B_1 &= b_1 e^{-x/\ell} .
\end{align}
The solutions for $B_0$ and $B_1$ are superpositions of four exponentials with
\begin{equation}
	\ell_{1,2}= \lambda \left(  1-  \frac{\omega_{\mathrm{dr}}^2}{2\omega_{\mathrm{pl}}^2} \pm \sqrt{ \frac{\omega_{\mathrm{dr}}^4}{4\omega_{\mathrm{pl}}^4} + \frac{M^2}{2}} \right)^{-1/2} ,~\ell_{3,4}= - \ell_{1,2} .
\end{equation}
While $\ell_1$ is generally real-valued and smaller than $\lambda$, $\ell_2$ is real-valued only for $\omega_{\mathrm{dr}} \lesssim \omega_{\mathrm{pl}}$ and imaginary for $\omega_{\mathrm{dr}} > \omega_{\mathrm{pl}}$. The absolute value of $\ell_2$ is larger than $\lambda$ for $\omega_{\mathrm{dr}} \sim \omega_{\mathrm{pl}}$. The solution for $B_2(x)$ is of the form
\begin{equation}
	B_2= b_2 e^{-x/\ell_0} ,
\end{equation}
where
\begin{equation}
	\ell_0= \pm \frac{\lambda}{\sqrt{1- \omega_{\mathrm{dr}}^2/\omega_{\mathrm{pl}}^2}}
\end{equation}
is real-valued for $\omega_{\mathrm{dr}} < \omega_{\mathrm{pl}}$ and imaginary for $\omega_{\mathrm{dr}} > \omega_{\mathrm{pl}}$.

In the red-detuned case of $\omega_{\mathrm{dr}} < \omega_{\mathrm{pl}}$, we exclude exponentially growing solutions and write the magnetic field inside the superconductor as
\begin{align} \label{eq:By_red}
	\begin{split}
		B_y^{\mathrm{(in)}} ={}& \beta_1 e^{-x/\ell_1} + \beta_2 e^{-x/\ell_2} + \frac{M \beta_1 e^{-x/\ell_1} \cos(\omega_{\mathrm{dr}}t)}{\omega_{\mathrm{dr}}^2/2\omega_{\mathrm{pl}}^2 + \sqrt{ \omega_{\mathrm{dr}}^4/4\omega_{\mathrm{pl}}^4 + M^2/2}} \\
		&+ \frac{M \beta_2 e^{-x/\ell_2} \cos(\omega_{\mathrm{dr}}t)}{\omega_{\mathrm{dr}}^2/2\omega_{\mathrm{pl}}^2 - \sqrt{ \omega_{\mathrm{dr}}^4/4\omega_{\mathrm{pl}}^4 + M^2/2}} + \beta_3 e^{-x/\ell_0} \sin(\omega_{\mathrm{dr}} t) .
	\end{split}
\end{align}
The corresponding electric field is
\begin{align} \label{eq:Ez_red}
	\begin{split}
		E_z^{\mathrm{(in)}} = \int \partial_t B_y^{\mathrm{(in)}} \mathrm{d}x={}& \frac{\ell_1 \omega_{\mathrm{dr}} M \beta_1 e^{-x/\ell_1} \sin(\omega_{\mathrm{dr}}t)}{\omega_{\mathrm{dr}}^2/2\omega_{\mathrm{pl}}^2 + \sqrt{ \omega_{\mathrm{dr}}^4/4\omega_{\mathrm{pl}}^4 + M^2/2}} + \frac{\ell_2 \omega_{\mathrm{dr}} M \beta_2 e^{-x/\ell_2} \sin(\omega_{\mathrm{dr}}t)}{\omega_{\mathrm{dr}}^2/2\omega_{\mathrm{pl}}^2 - \sqrt{ \omega_{\mathrm{dr}}^4/4\omega_{\mathrm{pl}}^4 + M^2/2}} \\
		&- \ell_0 \omega_{\mathrm{dr}} \beta_3 e^{-x/\ell_0} \cos(\omega_{\mathrm{dr}} t) .
	\end{split}
\end{align}
We note that red-detuned parametric driving induces an AC contribution to the magnetic field, which is less effectively screened than the DC magnetic field. For blue-detuned parametric driving, the induced AC part of the magnetic field leads to the formation of two standing waves.
As these standing waves are induced at the surface, we use the ansatz
\begin{align} \label{eq:By_blue}
	\begin{split}
		B_y^{\mathrm{(in)}} ={}& \beta_1 e^{-x/\ell_1} + \beta_2 \cos(x/|\ell_2|) + \frac{M \beta_1 e^{-x/\ell_1} \cos(\omega_{\mathrm{dr}}t)}{\omega_{\mathrm{dr}}^2/2\omega_{\mathrm{pl}}^2 + \sqrt{ \omega_{\mathrm{dr}}^4/4\omega_{\mathrm{pl}}^4 + M^2/2}} \\
		&+ \frac{M \beta_2 \cos(x/|\ell_2|) \cos(\omega_{\mathrm{dr}}t)}{\omega_{\mathrm{dr}}^2/2\omega_{\mathrm{pl}}^2 - \sqrt{ \omega_{\mathrm{dr}}^4/4\omega_{\mathrm{pl}}^4 + M^2/2}} + \beta_3 \cos(x/|\ell_0|) \sin(\omega_{\mathrm{dr}} t)
	\end{split}
\end{align}
for $\omega_{\mathrm{dr}} > \omega_{\mathrm{pl}}$. The electric field has the form
\begin{align} \label{eq:Ez_blue}
	\begin{split}
		E_z^{\mathrm{(in)}} = \int \partial_t B_y^{\mathrm{(in)}} \mathrm{d}x={}& \frac{\ell_1 \omega_{\mathrm{dr}} M \beta_1 e^{-x/\ell_1} \sin(\omega_{\mathrm{dr}}t)}{\omega_{\mathrm{dr}}^2/2\omega_{\mathrm{pl}}^2 + \sqrt{ \omega_{\mathrm{dr}}^4/4\omega_{\mathrm{pl}}^4 + M^2/2}} - \frac{|\ell_2| \omega_{\mathrm{dr}} M \beta_2 \sin(x/|\ell_2|) \sin(\omega_{\mathrm{dr}}t)}{\omega_{\mathrm{dr}}^2/2\omega_{\mathrm{pl}}^2 - \sqrt{ \omega_{\mathrm{dr}}^4/4\omega_{\mathrm{pl}}^4 + M^2/2}} \\
		&+ |\ell_0| \omega_{\mathrm{dr}} \beta_3 \sin(x/|\ell_0|) \cos(\omega_{\mathrm{dr}} t) .
	\end{split}
\end{align}

In general, parametric driving of a superconductor in the presence of a magnetic field causes emission of electromagnetic waves. Here, we consider the emission of electromagnetic waves from the left edge of the sample,
\begin{align}
	B_y^{\mathrm{(out)}} &= B_{\mathrm{ext}} + \alpha_1 \cos \bigl(\omega_{\mathrm{dr}} (t + x/c) \bigr) + \alpha_2 \sin \bigl(\omega_{\mathrm{dr}} (t + x/c) \bigr) , \\
	E_z^{\mathrm{(out)}} &= c \alpha_1 \cos \bigl(\omega_{\mathrm{dr}} (t + x/c) \bigr) + c \alpha_2 \sin \bigl(\omega_{\mathrm{dr}} (t + x/c) \bigr) .
\end{align}
Using the continuity of $B_y(x,t)$ and $E_z(x,t)$ at the surface of the sample, we determine the coefficients $\beta_1$, $\beta_2$ and $\beta_3$; see supplementary material for details of the calculation. In the red-detuned case, we obtain
\begin{align}
	\beta_1 &= \frac{B_{\mathrm{ext}}}{\zeta} \left(1+ \frac{\ell_2 \ell_0 \omega_{\mathrm{dr}}^2}{c^2} \right) \left( \frac{\omega_{\mathrm{dr}}^2}{2\omega_{\mathrm{pl}}^2} + \sqrt{ \frac{\omega_{\mathrm{dr}}^4}{4\omega_{\mathrm{pl}}^4} + \frac{M^2}{2}} \right) , \\
	\beta_2 &= - \frac{B_{\mathrm{ext}}}{\zeta} \left(1+ \frac{\ell_1 \ell_0 \omega_{\mathrm{dr}}^2}{c^2} \right) \left( \frac{\omega_{\mathrm{dr}}^2}{2\omega_{\mathrm{pl}}^2} - \sqrt{ \frac{\omega_{\mathrm{dr}}^4}{4\omega_{\mathrm{pl}}^4} + \frac{M^2}{2}} \right) , \\
	\beta_3 &= \frac{\omega_{\mathrm{dr}} M B_{\mathrm{ext}}}{\zeta c} \left(\ell_1 - \ell_2 \right) ,
\end{align}
where
\begin{align}
		\zeta= \left(1+ \frac{\ell_2 \ell_0 \omega_{\mathrm{dr}}^2}{c^2} \right) \left( \frac{\omega_{\mathrm{dr}}^2}{2\omega_{\mathrm{pl}}^2} + \sqrt{ \frac{\omega_{\mathrm{dr}}^4}{4\omega_{\mathrm{pl}}^4} + \frac{M^2}{2}} \right) - \left(1+ \frac{\ell_1 \ell_0 \omega_{\mathrm{dr}}^2}{c^2} \right) \left( \frac{\omega_{\mathrm{dr}}^2}{2\omega_{\mathrm{pl}}^2} - \sqrt{ \frac{\omega_{\mathrm{dr}}^4}{4\omega_{\mathrm{pl}}^4} + \frac{M^2}{2}} \right) .
\end{align}
In the blue-detuned case, we find
\begin{align}
	\beta_1 &= \frac{B_{\mathrm{ext}}}{2} \left(1+ \frac{\omega_{\mathrm{dr}}^2/2\omega_{\mathrm{pl}}^2}{\sqrt{ \omega_{\mathrm{dr}}^4/4\omega_{\mathrm{pl}}^4 + M^2/2}} \right) , \\
	\beta_2 &= \frac{B_{\mathrm{ext}}}{2} \left(1- \frac{\omega_{\mathrm{dr}}^2/2\omega_{\mathrm{pl}}^2}{\sqrt{ \omega_{\mathrm{dr}}^4/4\omega_{\mathrm{pl}}^4 + M^2/2}} \right) , \\
	\beta_3 &= \frac{B_{\mathrm{ext}}}{2c} \frac{\ell_1 \omega_{\mathrm{dr}} M}{\sqrt{ \omega_{\mathrm{dr}}^4/4\omega_{\mathrm{pl}}^4 + M^2/2}} .
\end{align}

\begin{figure}[!t]
	\centering
	\includegraphics[scale=1]{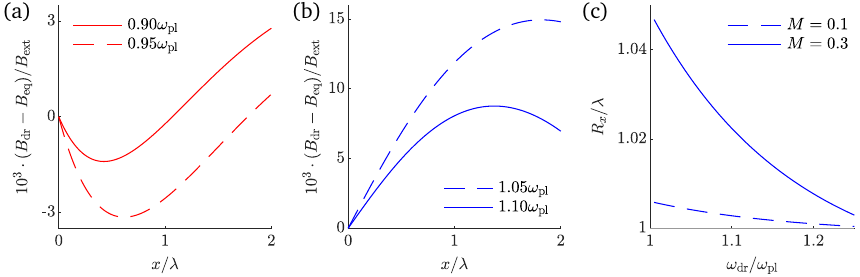}
	\caption{Analytical results for the Meissner screening in a parametrically driven superconductor. (a) Spatial dependence of the DC magnetic field for $M=0.3$ and two different red-detuned driving frequencies. (b) Spatial dependence of the DC magnetic field for $M=0.3$ and two different blue-detuned driving frequencies. (c) Attenuation length $R_x$ of the DC magnetic field as a function of the driving frequency for two different driving strengths. In panels (a) and (b), the equilibrium magnetic field $B_{\mathrm{eq}}= B_{\mathrm{ext}} \exp(- x/\lambda)$ is subtracted, where $\lambda$ is the London penetration depth.}
	\label{fig:fig3}
\end{figure}

In Figs.~\ref{fig:fig3}(a) and \ref{fig:fig3}(b), we show how red- and blue-detuned driving modifies the spatial dependence of the DC magnetic field inside the superconductor. In the red-detuned case, the DC part of the magnetic field is the sum of two exponentially decaying contributions. As the driving frequency approaches the plasma frequency, the length scale of the first decay converges to a value below the equilibrium penetration depth,
\begin{equation}
	\ell_1 \rightarrow \frac{\lambda}{\sqrt{1+ M^2/2}} .
\end{equation}

This feature of the response by itself indicates a parametric enhancement of the Meissner screening.
However, the length scale $\ell_2$ of the second exponential decay is generally larger than the equilibrium penetration depth. Thus, the enhancement of the Meissner screening is lessened or reverted. While $\ell_2$ diverges for $\omega_{\mathrm{dr}} \rightarrow \omega_{\mathrm{pl}}$, the prefactor of the second exponential decay vanishes in this limit. Taking both into account, the Meissner screening is enhanced as the driving frequency is slightly below the plasma frequency.  For larger detuning, the enhanced screening is effective only on a short length scale. Further away from the surface, the slower decaying contribution dominates such that the DC magnetic field is larger than in the absence of driving. This is visible in Fig.~\ref{fig:fig3}(a).

For blue-detuned driving, the DC part of the magnetic field is the sum of a contribution that decays exponentially on the length scale $\ell_1 < \lambda$ and a spatially oscillating contribution. As evidenced by Fig.~\ref{fig:fig3}(b), the spatially oscillating contribution reduces the Meissner screening such that the DC magnetic field is larger than in the absence of driving. 
In the supplementary material, we present the spatial dependence of the DC magnetic field explicitly, using a higher driving amplitude of $M=0.6$.

Figure~\ref{fig:fig3}(c) displays the attenuation length of the magnetic field as a function of the driving frequency for blue-detuned driving. The attenuation length is defined by the condition $B_y(x=R_x)= B_{\mathrm{ext}} \exp(-1)$. While the attenuation length equals the London penetration depth in equilibrium, it is increased by slightly blue-detuned driving. The attenuation length grows as the detuning of the driving frequency from the plasma frequency is decreased and the driving strength is increased.

In our simulations, we apply a static magnetic field at the surface of the superconductor. The analytical solution for this boundary condition is provided in the supplementary material. We find that the solution for the DC magnetic field inside the superconductor is not affected in the case of blue-detuned driving. However, the modified boundary condition suppresses the enhancement regime for red-detuned driving.

\subsection{Numerical results for an isotropic superconductor}
\label{sec:numerical_iso}
\begin{figure}[!b]
	\centering
	\includegraphics[scale=1]{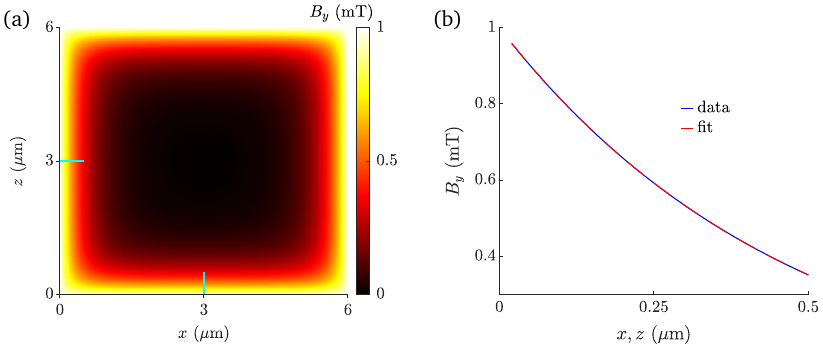}
	\caption{Expulsion of a static magnetic field from an isotropic superconductor in equilibrium. (a) Spatial dependence of the magnetic field. (b) Exponential fit to the decay of the magnetic field along the cyan lines in (a). The decay of the magnetic field is the same for both lines. The fitted value of the London penetration depth $\lambda_{\mathrm{eq}}= 478~\mathrm{nm}$ is in good agreement with the analytical prediction of 477 nm. The sample parameters are the same as in Fig.~\ref{fig:fig2}, except for the sample size.}
	\label{fig:fig4}
\end{figure}

To simulate the Meissner effect, we apply a small surface magnetic field along the $y$ axis, i.e., $\mathbf{B}= B_{\mathrm{ext}} \mathbf{\hat{y}}$. Throughout this paper, we use $B_{\mathrm{ext}}= 1~\mathrm{mT}$. Note that we obtain consistent results for $B_{\mathrm{ext}}= 0.1~\mathrm{mT}$, which confirms that the linear response is measured.
In Fig.~\ref{fig:fig4}, we present equilibrium results for an isotropic superconductor with the same parameters as in Section~\ref{sec:Conductivity}, except for the sample size. The plasma frequency is $\omega_{\mathrm{pl}}/2\pi= 100~\mathrm{THz}$ and the edge length is $L= 6~\mathrm{\mu m}$ along both axes. We see in Fig.~\ref{fig:fig4}(a) that the magnetic field $B_y(x,z)$ is screened away from the surface, which is the characteristic response of a superconductor to a magnetic field. As shown in Fig.~\ref{fig:fig4}(b), the decay of the magnetic field from the sample surfaces is well captured by the exponential fit functions $B_y (x,0)= B_{\mathrm{ext}} \, \mathrm{exp} \left( -x/\lambda_{\mathrm{eq}} \right)$ and $B_y (0,z)= B_{\mathrm{ext}} \, \mathrm{exp} \left( -z/\lambda_{\mathrm{eq}} \right)$, with $\lambda_{\mathrm{eq}}$ being the only free parameter. The fitted value $\lambda_{\mathrm{eq}}= 478~\mathrm{nm}$ of the London penetration depth is in excellent agreement with the analytical prediction $\lambda= c/\omega_{\mathrm{pl}}= 477~\mathrm{nm}$. In fact, the numerical value converges to the analytical prediction for larger sample size; see supplementary material.

In the remainder of this section, we investigate the response of an isotropic superconductor to an external magnetic field in the presence of parametric driving as defined in Eq.~\eqref{eq:driving}. We characterize the Meissner screening in the driven state by the attenuation lengths $R_x$ and $R_z$. The attenuation length $R_x$ quantifies the Meissner screening at the center of the left sample surface, i.e., for $x \ll L_x$ and $z= L_z/2$. To resolve changes of $R_x$ below the discretization length $d$, we interpolate the magnetic field linearly between the plaquettes left and right of $x= R_x$. The attenuation length along the $z$ axis, $R_z$, is determined analogously. In equilibrium, the attenuation lengths equal the London penetration depth, i.e., $R_x= R_z= \lambda_{\mathrm{eq}}$.

\begin{figure}[!t]
	\centering
	\includegraphics[scale=1]{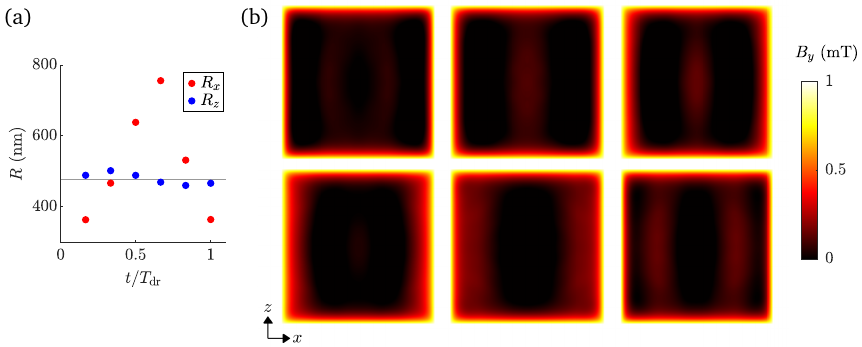}
	\caption{Expulsion of a static magnetic field from an isotropic superconductor in the presence of blue-detuned driving of the $z$-axis tunneling. (a) Dynamics of the attenuation lengths during one driving cycle $T_{\mathrm{dr}}= 2\pi/\omega_{\mathrm{dr}}$. The gray line indicates the equilibrium value $\lambda_{\mathrm{eq}}= 477~\mathrm{nm}$. (b) Spatial dependence of the magnetic field during one driving cycle, corresponding to the attenuation lengths in (a). The snapshots are ordered from left to right and top to bottom, with the upper left snapshot taken at $T_{\mathrm{dr}}/6$. The driving frequency is $\omega_{\mathrm{dr}}= 1.1 \omega_{\mathrm{pl}}$ and the driving amplitude is $M=0.3$. The sample size is $12 \times 12~\mathrm{\mu m^2}$.}
	\label{fig:fig5}
\end{figure}

We consider a superconducting sample with the same parameters as before but with a sample size of $12 \times 12~\mathrm{\mu m^2}$ to ensure the convergence of our results.
First, we choose the driving frequency $\omega_{\mathrm{dr}}/2\pi= 110~\mathrm{THz}$ and the driving amplitude $M=0.3$, consistent with the conductivity measurements shown in Fig.~\ref{fig:fig2}. Once a steady state is reached, the magnetic field inside the superconductor oscillates with the driving frequency. Snapshots of the time evolution of the magnetic field during one driving cycle are displayed in Fig.~\ref{fig:fig5}(b). The parametric driving with $\omega_{\mathrm{dr}} > \omega_{\mathrm{pl}}$ has two main effects. Firstly, electromagnetic waves generated at the left and right surfaces are transmitted into the bulk of the sample. However, the magnitude of the magnetic field inside the superconductor is strongly suppressed compared to the surface field $B_{\mathrm{ext}}$. Secondly, the attenuation lengths are no longer isotropic and exhibit an oscillatory behavior in time. As evidenced by Fig.~\ref{fig:fig5}(a), $R_x$ exhibits a pronounced oscillation, while $R_z$ has a small oscillation amplitude. Remarkably, we find that there is no generation of electromagnetic waves for driving frequencies red-detuned from the plasma frequency. The time evolution of the magnetic field for one example of red-detuned driving is shown in the supplementary material.

Next, we average the magnetic field over 1~ps with a detection rate of 5~PHz and evaluate the attenuation lengths $R_x= 488~\mathrm{nm}$ and $R_z= 479~\mathrm{nm}$ of the time-averaged magnetic field. The attenuation lengths in the driven state are both larger than the equilibrium value $\lambda_{\mathrm{eq}}= 477~\mathrm{nm} $ of this sample. So, while the parametric driving leads to a significant enhancement of $z$-axis transport, the time-averaged screening of magnetic fields is slightly reduced. This result also holds for larger sample size and AC magnetic fields with small frequencies $\sim$1~THz; see supplementary material. We note that the relative phase between the oscillation of $R_x$ and the oscillation of $R_z$ depends on the lateral sample size. This suggests that the modulation of $R_z$ is due to the transmission of electromagnetic waves from the left and right surfaces.

\begin{figure}[!t]
	\centering
	\includegraphics[scale=1]{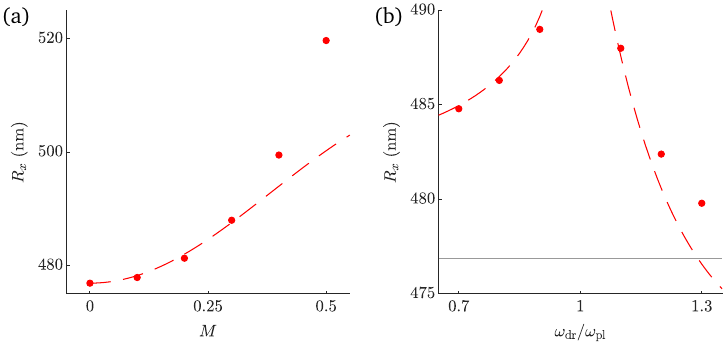}
	\caption{Attenuation length $R_x$ of the time-averaged magnetic field for different driving strengths and frequencies. (a) $R_x$ for $\omega_{\mathrm{dr}}= 1.1 \omega_{\mathrm{pl}}$ and different driving strengths. (b) $R_x$ for $M=0.3$ and different driving frequencies. The gray line indicates the equilibrium value $\lambda_{\mathrm{eq}}= 477~\mathrm{nm}$. In both panels, the dashed red line indicates the analytical solution for $R_x$. The sample size is $12 \times 12~\mathrm{\mu m^2}$, except for the two largest driving frequencies in (b), where converged results are obtained for a sample size of $16 \times 16~\mathrm{\mu m^2}$.}
	\label{fig:fig6}
\end{figure}

We proceed by varying the driving strength and the driving frequency. Figure~\ref{fig:fig6}(a) demonstrates that the attenuation length $R_x$ grows monotonically with increasing driving amplitude. The data points in Fig.~\ref{fig:fig6}(b) indicate a divergence of $R_x$ as the driving frequency approaches the plasma frequency. We compare our numerical results to the analytical solution from Section~\ref{sec:analytical}, applying the boundary condition of a static magnetic field at the sample surface. The numerical results show good agreement with the analytical solution, except for the data point at $\omega_{\mathrm{dr}}= 1.3 \omega_{\mathrm{pl}}$. This discrepancy is due to the approximations that we used in the derivation of the analytical solution. For example, we neglected the spatial dependence of the magnetic field along the $z$ axis and temporal oscillations at higher harmonics of the driving frequency. Thus, the analytical solution is valid only close to the sample surface and does not capture the propagation of electromagnetic waves in the case of blue-detuned driving.

\subsection{Numerical results for an anisotropic superconductor}
\label{sec:numerical_aniso}
In this section, we study the effect of parametric driving on the magnetic response of an anisotropic superconductor. Since our analytical arguments in Section~\ref{sec:analytical} are not limited to an isotropic superconductor, we expect a similar reduction of the Meissner screening for an anisotropic superconductor.
In cuprate superconductors, the ratio between the in-plane plasma frequency and the (lower) $c$-axis plasma frequency is of the order of 100. Due to numerical constraints, we choose plasma frequencies with a smaller ratio. In the following, we consider a superconductor with the plasma frequencies $\omega_x/2\pi= 300~\mathrm{THz}$ and $\omega_z/2\pi= 50~\mathrm{THz}$ along the $x$ axis and the $z$ axis, respectively. Consistent with the relations $\lambda_x= c/\omega_x$ and $\lambda_z= c/\omega_z$, we find the attenuation lengths $R_x= \lambda_z= 954~\mathrm{nm}$ and $R_z= \lambda_x= 159~\mathrm{nm}$ in equilibrium. The sample size is $24 \times 6~\mathrm{\mu m^2}$.

\begin{figure}[!t]
	\centering
	\includegraphics[scale=1]{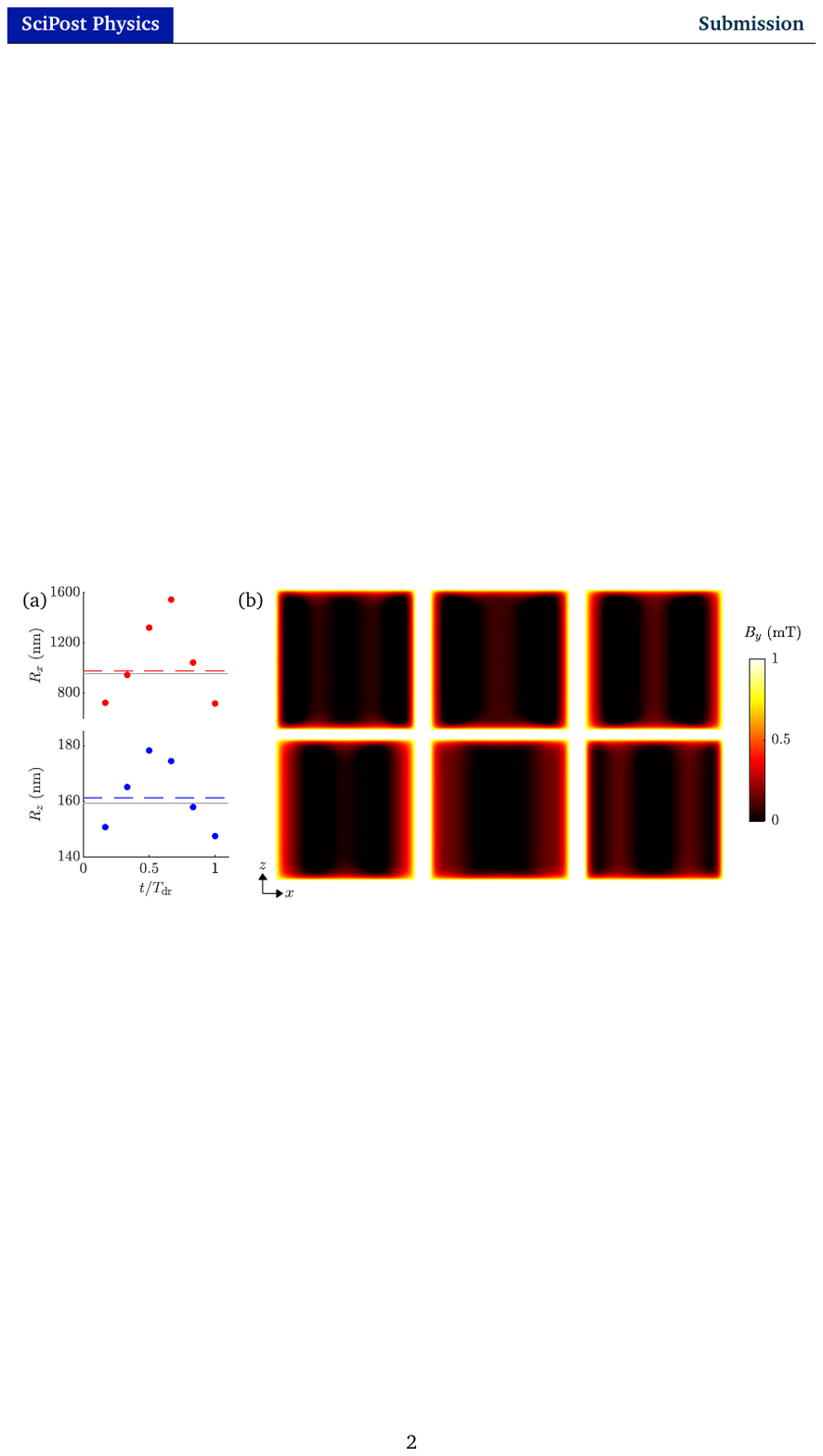}
	\caption{Expulsion of a static magnetic field from an anisotropic superconductor in the presence of blue-detuned driving of the $z$-axis tunneling. (a) Attenuation lengths as a function of time during one driving cycle $T_{\mathrm{dr}}= 2\pi/\omega_{\mathrm{dr}}$. The gray lines indicate the equilibrium values $\lambda_z$ and $\lambda_x$, and the dashed lines indicate the attenuation lengths of the time-averaged magnetic field in the driven state. (b) Spatial dependence of the magnetic field during one driving cycle, corresponding to the attenuation lengths in (a). The snapshots are ordered from left to right and top to bottom, with the upper left snapshot taken at $T_{\mathrm{dr}}/6$. The driving frequency is $\omega_{\mathrm{dr}}= 1.1 \omega_z$ and the driving amplitude is $M=0.3$. The sample size is $24 \times 6~\mathrm{\mu m^2}$.}
	\label{fig:fig7}
\end{figure}

We then add parametric driving with frequency $\omega_{\mathrm{dr}}/2\pi= 55~\mathrm{THz}$ and amplitude $M=0.3$.
We show in Fig.~\ref{fig:fig7}(b) that the time evolution of the magnetic field during one driving cycle is comparable to the isotropic case. However, the spatial patterns are sharper and more pronounced, especially towards the top and bottom of the sample. While the oscillation amplitude of $R_x$ compared to its temporal average is similar to the isotropic case, the oscillation amplitude of $R_z$ is significantly larger as shown in Fig.~\ref{fig:fig7}(a). This further indicates that the modulation of $R_z$ is a consequence of electromagnetic waves transmitted into the bulk. Here, the attenuation lengths of the time-averaged magnetic field are $R_x= 977~\mathrm{nm}$ and $R_z= 161~\mathrm{nm}$ in the driven state. The increase of $R_x$ by approximately 2\% is in good agreement with our observation for an isotropic superconductor, where we also used $\omega_{\mathrm{dr}}= 1.1 \omega_z$ and $M=0.3$. The increase of $R_z$ by more than 1\% is considerably larger than in the isotropic case.

\section{Conclusion}
\label{sec:Conclusion}
In conclusion, we have presented the response of light-driven superconductors to magnetic fields for the scenario of parametrically driven $z$-axis tunneling. For driving with a frequency blue-detuned from the plasma frequency, we find an enhancement of $z$-axis transport and a reduction of the Meissner screening along the $x$~axis, the direction perpendicular to the parametric drive and the applied magnetic field. This key result is in contrast to the equilibrium behavior of superconductors. In the absence of driving, both London theory \cite{London1935} and our model in Eq.~\eqref{eq:Lagrangian} predict that an enhancement of the low-frequency $\sigma_2(\omega)$ along the $z$~axis implies an enhancement of the Meissner screening along the $x$~axis. Our simulations demonstrate the breakdown of this general relation in a driven superconductor. In fact, the screening of DC magnetic fields is reduced for slightly blue-detuned driving, which can be understood analytically based on a minimal model that we derived in this work. According to our analytical calculations, the screening of DC magnetic fields shows a tendency to be enhanced for slightly red-detuned driving. This enhanced screening is enabled by emission of electromagnetic waves from the superconductor. If the emission of electromagnetic waves is suppressed, as in our simulations, the Meissner screening for red-detuned driving is generally less effective than in the absence of driving. We emphasize that we observe similar behavior for isotropic and anisotropic superconductors.

Our findings suggest that the enhancement of the low-frequency conductivity is naturally accompanied by a suppression of the Meissner effect in the parametrically driven scenario that we consider. The parametric driving mixes the Josephson plasmon into the low-frequency response. While this admixture provides an enhanced conductivity, it results in a suppressed Meissner screening due to the transmission of unscreened plasma excitations into the superconductor. More generally, our results suggest that the light-induced state is a genuinely non-equilibrium state, rather than a renormalized equilibrium state, in which some of the reasoning derived from equilibrium superconductors does not apply.

Our work is relevant for the interpretation of pump-probe experiments on light-driven superconductors, particularly cuprates. An improved understanding of these experiments might eventually provide new insights into the nature of the superconducting state in unconventional superconductors.

\begin{acknowledgments}
We thank Gregor Jotzu, Lukas Broers and Jim Skulte for stimulating discussions.
This work is supported by the Deutsche Forschungsgemeinschaft (DFG) in the framework of SFB~925, Project No.~170620586, and the Cluster of Excellence ``Advanced Imaging of Matter" (EXC~2056), Project No.~390715994.
\end{acknowledgments}

\bibliography{biblio}

\end{document}


\title{Supplementary material for\\Parametric control of Meissner screening in light-driven superconductors}

\author{Guido Homann}
\affiliation{Zentrum f\"ur Optische Quantentechnologien and Institut f\"ur Laserphysik, 
	Universit\"at Hamburg, 22761 Hamburg, Germany}

\author{Jayson G. Cosme}
\affiliation{National Institute of Physics, University of the Philippines, Diliman, Quezon City 1101, Philippines}

\author{Ludwig Mathey}
\affiliation{Zentrum f\"ur Optische Quantentechnologien and Institut f\"ur Laserphysik, 
	Universit\"at Hamburg, 22761 Hamburg, Germany}
\affiliation{The Hamburg Centre for Ultrafast Imaging, Luruper Chaussee 149, 22761 Hamburg, Germany}

\maketitle
\tableofcontents

\clearpage
\section{Redistribution of spectral weight}
\label{sec:spectralWeight}
\begin{figure}[!h]
	\centering
	\includegraphics[scale=1]{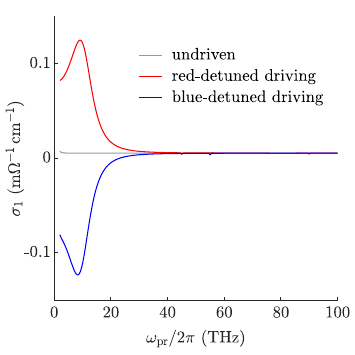}
	\caption{Real part of the optical conductivity of a parametrically driven superconductor. The driving amplitude is $M=0.3$. The driving frequencies are $\omega_{\mathrm{dr}}= 0.9 \omega_{\mathrm{pl}}$ and $\omega_{\mathrm{dr}}= 1.1 \omega_{\mathrm{pl}}$, consistent with Fig.~2 in the main text.}
	\label{fig:fig1}
\end{figure}

As shown in Fig.~\ref{fig:fig1}, the real part $\sigma_1$ of the optical conductivity exhibits a maximum around $\omega^*= |\omega_{\mathrm{pl}} - \omega_{\mathrm{dr}}|$ for red-detuned driving, while it exhibits a minimum around $\omega^*$ for blue-detuned driving. These extrema of $\sigma_1$ correspond to parametric attenuation/amplification as discussed in Ref.~\cite{Homann2022}.
Following the notation in Ref.~\cite{Resta2018}, we write the optical conductivity as
\begin{equation}
	\sigma (\omega)= D \left(\delta(\omega) + \frac{i}{\pi \omega} \right) + \sigma^{\mathrm{regular}} (\omega) .
\end{equation}
The total spectral weight is then given by
\begin{equation}
	S= \int_{0}^{\infty} \mathrm{d}\omega \, \sigma_1(\omega)= \frac{D}{2} + \int_{0}^\infty \mathrm{d}\omega \, \sigma_1^{\mathrm{regular}} (\omega) .
\end{equation}
Here, we evaluate the spectral weight difference
\begin{equation}
	\Delta W= \int_{\omega_a}^{\omega_b} \mathrm{d}\omega \left( \sigma_1^{\mathrm{driven}} (\omega) - \sigma_1^{\mathrm{undriven}} (\omega) \right)
\end{equation}
with $\omega_a/2\pi= 2~\mathrm{THz}$ and $\omega_b/2\pi= 100~\mathrm{THz}$. For red-detuned driving, we find $\Delta W_{\mathrm{red}} \approx -0.84 \cdot (D_{\mathrm{red}} - D_0)/2$, while we obtain $\Delta W_{\mathrm{blue}} \approx -0.94 \cdot (D_{\mathrm{blue}} - D_0)/2$ for blue-detuned driving. Note that the absolute value of $\Delta W$ is underestimated in both cases due to the lower cutoff at $\omega_a$.
These results indicate that the parametric driving redistributes spectral weight, while the total spectral weight is conserved.

\section{Details on the analytical solution}
\label{sec:analyticalDetails}

\subsection{Emitting boundary condition}
\label{sec:emittingBC}
\begin{figure}[!h]
	\centering
	\includegraphics[scale=1]{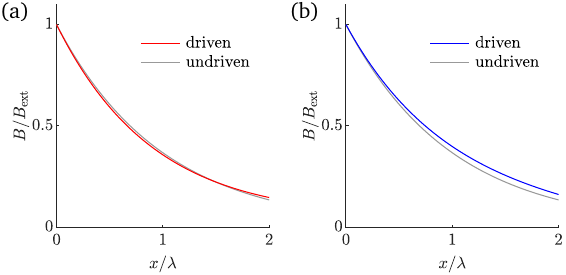}
	\caption{Spatial dependence of the DC magnetic field for M = 0.6. (a) Red-detuned driving, $\omega_{\mathrm{dr}}=0.95 \omega_{\mathrm{pl}}$. (b) Blue-detuned driving, $\omega_{\mathrm{dr}}=1.05 \omega_{\mathrm{pl}}$.}
	\label{fig:fig2}
\end{figure}

As mentioned in Section~4.1 of the main text, the electromagnetic waves emitted by the parametrically driven superconductor have the form
\begin{align}
	B_y^{\mathrm{(out)}} &= B_{\mathrm{ext}} + \alpha_1 \cos \bigl(\omega_{\mathrm{dr}} (t + x/c) \bigr) + \alpha_2 \sin \bigl(\omega_{\mathrm{dr}} (t + x/c) \bigr) , \\
	E_z^{\mathrm{(out)}} &= c \alpha_1 \cos \bigl(\omega_{\mathrm{dr}} (t + x/c) \bigr) + c \alpha_2 \sin \bigl(\omega_{\mathrm{dr}} (t + x/c) \bigr) .
\end{align}
For the magnetic and the electric field inside the superconductor, we use the general ansatz from Section~4.1 of the main text.
In the red-detuned case, the continuity of $B_y(x,t)$ at $x=0$ implies
\begin{align}
	B_{\mathrm{ext}} &= \beta_1 + \beta_2  , \\
	\alpha_1 &= \frac{M \beta_1}{\omega_{\mathrm{dr}}^2/2\omega_{\mathrm{pl}}^2 + \sqrt{ \omega_{\mathrm{dr}}^4/4\omega_{\mathrm{pl}}^4 + M^2/2}} + \frac{M \beta_2}{\omega_{\mathrm{dr}}^2/2\omega_{\mathrm{pl}}^2 - \sqrt{ \omega_{\mathrm{dr}}^4/4\omega_{\mathrm{pl}}^4 + M^2/2}} , \\
	\alpha_2 &= \beta_3 .
\end{align}
The continuity of $E_z(x,t)$ at $x=0$ implies
\begin{align}
	c \alpha_1 &= -\ell_0 \omega_{\mathrm{dr}} \beta_3 , \\
	c \alpha_2 &= \frac{\ell_1 \omega_{\mathrm{dr}} M \beta_1}{\omega_{\mathrm{dr}}^2/2\omega_{\mathrm{pl}}^2 + \sqrt{ \omega_{\mathrm{dr}}^4/4\omega_{\mathrm{pl}}^4 + M^2/2}} + \frac{\ell_2 \omega_{\mathrm{dr}} M \beta_2}{\omega_{\mathrm{dr}}^2/2\omega_{\mathrm{pl}}^2 - \sqrt{ \omega_{\mathrm{dr}}^4/4\omega_{\mathrm{pl}}^4 + M^2/2}} .
\end{align}
Thus, we obtain
\begin{align}
	\beta_1 &= \frac{B_{\mathrm{ext}}}{\zeta} \left(1+ \frac{\ell_2 \ell_0 \omega_{\mathrm{dr}}^2}{c^2} \right) \left( \frac{\omega_{\mathrm{dr}}^2}{2\omega_{\mathrm{pl}}^2} + \sqrt{ \frac{\omega_{\mathrm{dr}}^4}{4\omega_{\mathrm{pl}}^4} + \frac{M^2}{2}} \right) , \\
	\beta_2 &= - \frac{B_{\mathrm{ext}}}{\zeta} \left(1+ \frac{\ell_1 \ell_0 \omega_{\mathrm{dr}}^2}{c^2} \right) \left( \frac{\omega_{\mathrm{dr}}^2}{2\omega_{\mathrm{pl}}^2} - \sqrt{ \frac{\omega_{\mathrm{dr}}^4}{4\omega_{\mathrm{pl}}^4} + \frac{M^2}{2}} \right) , \\
	\beta_3 &= \frac{\omega_{\mathrm{dr}} M B_{\mathrm{ext}}}{\zeta c} \left(\ell_1 - \ell_2 \right) ,
\end{align}
where
\begin{align}
	\zeta= \left(1+ \frac{\ell_2 \ell_0 \omega_{\mathrm{dr}}^2}{c^2} \right) \left( \frac{\omega_{\mathrm{dr}}^2}{2\omega_{\mathrm{pl}}^2} + \sqrt{ \frac{\omega_{\mathrm{dr}}^4}{4\omega_{\mathrm{pl}}^4} + \frac{M^2}{2}} \right) - \left(1+ \frac{\ell_1 \ell_0 \omega_{\mathrm{dr}}^2}{c^2} \right) \left( \frac{\omega_{\mathrm{dr}}^2}{2\omega_{\mathrm{pl}}^2} - \sqrt{ \frac{\omega_{\mathrm{dr}}^4}{4\omega_{\mathrm{pl}}^4} + \frac{M^2}{2}} \right) .
\end{align}
In the blue-detuned case, the continuity of $B_y(x,t)$ at $x=0$ implies
\begin{align}
	B_{\mathrm{ext}} &= \beta_1 + \beta_2  , \\
	\alpha_1 &= \frac{M \beta_1}{\omega_{\mathrm{dr}}^2/2\omega_{\mathrm{pl}}^2 + \sqrt{ \omega_{\mathrm{dr}}^4/4\omega_{\mathrm{pl}}^4 + M^2/2}} + \frac{M \beta_2}{\omega_{\mathrm{dr}}^2/2\omega_{\mathrm{pl}}^2 - \sqrt{ \omega_{\mathrm{dr}}^4/4\omega_{\mathrm{pl}}^4 + M^2/2}} , \\
	\alpha_2 &= \beta_3 .
\end{align}
The continuity of $E_z$ at $x=0$ implies
\begin{align}
	c \alpha_1 &= 0 , \\
	c \alpha_2 &= \frac{\ell_1 \omega_{\mathrm{dr}} M \beta_1}{\omega_{\mathrm{dr}}^2/2\omega_{\mathrm{pl}}^2 + \sqrt{ \omega_{\mathrm{dr}}^4/4\omega_{\mathrm{pl}}^4 + M^2/2}} .
\end{align}
We find
\begin{align}
	\beta_1 &= \frac{B_{\mathrm{ext}}}{2} \left(1+ \frac{\omega_{\mathrm{dr}}^2/2\omega_{\mathrm{pl}}^2}{\sqrt{ \omega_{\mathrm{dr}}^4/4\omega_{\mathrm{pl}}^4 + M^2/2}} \right) , \\
	\beta_2 &= \frac{B_{\mathrm{ext}}}{2} \left(1- \frac{\omega_{\mathrm{dr}}^2/2\omega_{\mathrm{pl}}^2}{\sqrt{ \omega_{\mathrm{dr}}^4/4\omega_{\mathrm{pl}}^4 + M^2/2}} \right) , \\
	\beta_3 &= \frac{\ell_1 \omega_{\mathrm{dr}} M B_{\mathrm{ext}}/c}{2 \sqrt{ \omega_{\mathrm{dr}}^4/4\omega_{\mathrm{pl}}^4 + M^2/2}} .
\end{align}
Figure~\ref{fig:fig2} displays the analytical solution for the spatial dependence of the DC magnetic field explicitly. Here, the driving amplitude is chosen relatively high such that the effect of the parametric driving is visible.

\subsection{Static boundary condition}
\label{sec:staticBC}
\begin{figure}[!h]
	\centering
	\includegraphics[scale=1]{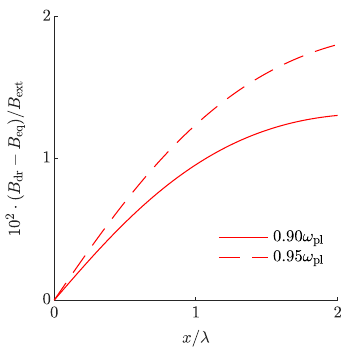}
	\caption{Spatial dependence of the DC magnetic field for $M=0.3$ and two different red-detuned driving frequencies, applying the boundary condition $B_y(x=0, t)= B_{\mathrm{ext}}$ to the analytical solution. The equilibrium magnetic field $B_{\mathrm{eq}}= B_{\mathrm{ext}} \exp(- x/\lambda)$ is subtracted, where $\lambda$ is the London penetration depth.}
	\label{fig:fig3}
\end{figure}

For comparison with the numerical simulations, we apply the boundary condition $B_y(x=0, t)= B_{\mathrm{ext}}$. With the general ansatz for the magnetic field inside the superconductor from Section~4.1 of the main text, we then obtain
\begin{align}
	\beta_1 &= \frac{B_{\mathrm{ext}}}{2} \left(1+ \frac{\omega_{\mathrm{dr}}^2/2\omega_{\mathrm{pl}}^2}{\sqrt{ \omega_{\mathrm{dr}}^4/4\omega_{\mathrm{pl}}^4 + M^2/2}} \right) , \\
	\beta_2 &= \frac{B_{\mathrm{ext}}}{2} \left(1- \frac{\omega_{\mathrm{dr}}^2/2\omega_{\mathrm{pl}}^2}{\sqrt{ \omega_{\mathrm{dr}}^4/4\omega_{\mathrm{pl}}^4 + M^2/2}} \right) , \\
	\beta_3 &= 0
\end{align}
for both red- and blue-detuned driving. In the case of blue-detuned driving, the DC part of the magnetic field is the same as for the emitting boundary condition considered in the previous section. For red-detuned driving, however, the modified boundary condition qualitatively changes the solution. In contrast to the emitting case, $\beta_2$ does not vanish in the limit of $\omega_{\mathrm{dr}} \rightarrow \omega_{\mathrm{pl}}$ such that the divergence of $\ell_2$ implies a less effective Meissner screening than in the absence of driving; see Fig.~\ref{fig:fig3}. Our analytical results are in good agreement with the numerical simulations as evidenced by Fig.~6 in the main text.

\section{Simulation parameters}
\label{sec:parameters}
The parameters of the simulated superconductors are summarized in Table~\ref{tab:parameters}. Our choice of $\mu$ and $g$ implies a Cooper pair density of $n_0= \mu/g = 2 \times 10^{21}~\mathrm{cm^{-3}}$ in equilibrium.
The discretization length $d$ is of the order of the Ginzburg-Landau estimate for the coherence length,
\begin{equation}
	\xi= \sqrt{\frac{t_{xy} d^2}{\mu}}= \sqrt{\frac{t_z d^2}{\mu}}=243~\text{\AA}
\end{equation}
for the isotropic sample. The Higgs frequency is
\begin{equation}
	\omega_{\mathrm{H}}= \sqrt{\frac{2\mu}{K \hbar^2}}= 2\pi \times 5~\mathrm{THz} .
\end{equation}

\begin{table}[!tb]
	\centering
	\caption{Model parameters of the simulated superconductors. If a second value is provided, it refers to the anisotropic sample.}
	\renewcommand{\arraystretch}{1.5}
	\begin{tabular}{lr}
		\hline
		$K~(\text{meV}^{-1})$ & $4.7 \times 10^{-5}$ \\
		$\mu~(\text{meV})$ & $1.0 \times 10^{-3}$ \\
		$g~(\text{meV} \, \text{\AA}^3)$ & 0.5 \\
		$d~(\text{\AA})$ & 200 \\
		\hline
		$\gamma_{\mathrm{sc}}/2\pi~(\mathrm{THz})$ & 1 \\
		$\gamma_{\mathrm{el},xy}/2\pi~(\mathrm{THz})$ & 10, 30 \\
		$\gamma_{\mathrm{el},z}/2\pi~(\mathrm{THz})$ & 10, 5 \\
		\hline
		$t_{xy}~(\text{meV})$ & $1.48 \times 10^{-3}$, $1.33 \times 10^{-2}$ \\
		$t_z~(\text{meV})$ & $\qquad 1.48 \times 10^{-3}$, $3.7 \times 10^{-4}$ \\
		\hline
	\end{tabular}
	\renewcommand{\arraystretch}{1}
	\label{tab:parameters}
\end{table}

\section{Finite size analysis}
\label{sec:finiteSize}
\begin{figure}[!t]
	\centering
	\includegraphics[scale=1]{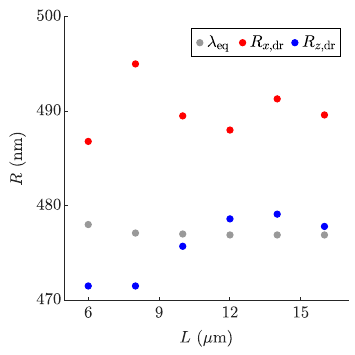}
	\caption{Attenuation lengths of the time-averaged magnetic field in an isotropic superconductor for different sample sizes. In equilibrium, the attenuation length equals the London penetration depth $\lambda_{\mathrm{eq}}$. The driving frequency is $\omega_{\mathrm{dr}}= 1.1 \omega_{\mathrm{pl}}$ and the driving amplitude is $M=0.3$.}
	\label{fig:fig4}
\end{figure}

In the main text, we present results for a parametrically driven superconductor with a sample size of $12 \times 12~\mathrm{\mu m^2}$. Here, we investigate the magnetic response of isotropic superconductors with different sample sizes $L \times L$. As visible in Fig.~\ref{fig:fig4}, the value of the equilibrium penetration depth is fully converged for samples with a size of $12 \times 12~\mathrm{\mu m^2}$ or larger. In fact, the numerical result agrees with the analytical prediction of $\lambda= 477~\mathrm{nm}$.

In the driven state, the values of the attenuation length $R_x$ exhibit no systematic drift for large samples, varying between 488 and 491~nm for $L \geq 12~\mathrm{\mu m}$. The values of the attenuation length $R_z$ vary between 478 and 479~nm for $L \geq 12~\mathrm{\mu m}$. We conclude that a sample with a size of $12 \times 12~\mathrm{\mu m^2}$ accurately reflects the magnetic response of large samples, both in equilibrium and in the presence of parametric driving.

\section{Time evolution of the magnetic field for red-detuned driving}
\label{sec:drivingFreq}
\begin{figure}[!t]
	\centering
	\includegraphics[scale=1]{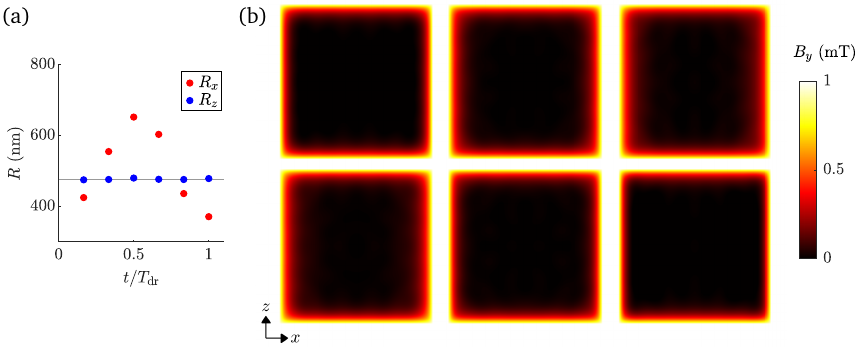}
	\caption{Expulsion of a static magnetic field from an isotropic superconductor in the presence of red-detuned driving of the $z$-axis tunneling. (a) Attenuation lengths as a function of time during one driving cycle $T_{\mathrm{dr}}= 2\pi/\omega_{\mathrm{dr}}$. The gray line indicates the equilibrium value $\lambda_{\mathrm{eq}}= 477~\mathrm{nm}$. (b) Spatial dependence of the magnetic field during one driving cycle, corresponding to the attenuation lengths in (a). The snapshots are ordered from left to right and top to bottom, with the upper left snapshot taken at $T_{\mathrm{dr}}/6$. The driving frequency is $\omega_{\mathrm{dr}}= 0.9 \omega_{\mathrm{pl}}$ and the driving amplitude is $M=0.3$. The sample size is $12 \times 12~\mathrm{\mu m^2}$.}
	\label{fig:fig5}
\end{figure}

In Fig.~\ref{fig:fig5}, we show the time evolution of the magnetic field for one example of parametric driving, in which the driving frequency is red-detuned from the plasma frequency. In stark contrast to the blue-detuned case shown in Fig.~5 in the main text, there is no visible transmission of electromagnetic waves from the left and right surfaces into the bulk of the sample. Thus, the modulation of $R_z$ is negligible, while the oscillation of $R_x$ is similar to the blue-detuned case.

\newpage
\section{AC Meissner effect in a parametrically driven superconductor}
\label{sec:acMeissner}
Finally, we briefly discuss the response of a parametrically driven isotropic superconductor to AC magnetic fields, i.e., $B_{\mathrm{ext}} \rightarrow B_{\mathrm{ext}} \cos(\omega_{\mathrm{pr}} t)$. Once a steady state is reached, we record the time evolution of the magnetic field for 10~ps with a detection rate of 5~PHz. We then compute the Fourier transform of the magnetic field to evaluate the component of $B_y(x,z)$ that oscillates with the probe frequency $\omega_{\mathrm{pr}}$. Eventually, we determine the attenuation lengths $R_x$ and $R_z$ as in the case of a static magnetic field.

\begin{figure}[!t]
	\centering
	\includegraphics[scale=1]{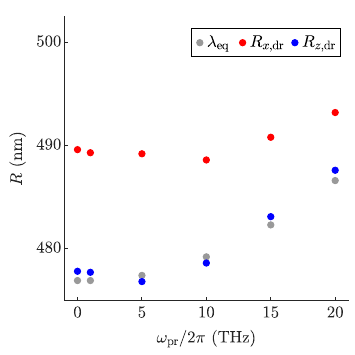}
	\caption{Attenuation lengths of the time-averaged magnetic field for different probe frequencies. In equilibrium, the attenuation length equals the London penetration depth $\lambda_{\mathrm{eq}}$. The driving frequency is $\omega_{\mathrm{dr}}= 1.1 \omega_{\mathrm{pl}}$ and the driving amplitude is $M=0.3$. The sample size is $16 \times 16~\mathrm{\mu m^2}$.}
	\label{fig:fig6}
\end{figure}

Figure~\ref{fig:fig6} displays the attenuation lengths as a function of the probe frequency. Importantly, the attenuation lengths of an AC magnetic field with $\omega_{\mathrm{pr}}/2\pi= 1~\mathrm{THz}$ approach the attenuation lengths of a static magnetic field. Therefore, it is sufficient to probe the sample with a static magnetic field in order to obtain the low-frequency limit of its magnetic response. This is particularly relevant when comparing the magnetic response to the optical response in Section~3 of the main text. For increasing probe frequency, the equilibrium penetration depth also increases, which can be understood from an analytical perspective. As mentioned in Section~4.1 of the main text, Maxwell's equations imply
\begin{equation} \label{eq:combiMaxwell}
	\frac{1}{\mu_0} \left( \frac{1}{c^2} \partial_t^2 - \mathbf{\nabla}^2 \right) \mathbf{B} = \mathbf{\nabla} \times \mathbf{J} .
\end{equation}
We insert the London equation $\mathbf{\nabla} \times \mathbf{J}= - \mathbf{B}/(\mu_0 \lambda^2)$ into Eq.~\eqref{eq:combiMaxwell}, where $\lambda$ is the London penetration depth of a static magnetic field. This leads to
\begin{equation}
	- \left( \frac{1}{c^2} \partial_t^2 - \mathbf{\nabla}^2 \right) \mathbf{B}= \frac{1}{\lambda^2} \mathbf{B} ,
\end{equation}
implying
\begin{equation} \label{eq:lambda_ac}
	\lambda(\omega_{\mathrm{pr}})= \frac{\lambda}{\sqrt{1 - \omega_{\mathrm{pr}}^2/\omega_{\mathrm{pl}}^2}} .
\end{equation}
For example, the numerical value of $\lambda_{\mathrm{eq}}= 487~\mathrm{nm}$ for $\omega_{\mathrm{pr}}/2\pi= 20~\mathrm{THz}$ agrees with the analytical prediction based on Eq.~\eqref{eq:lambda_ac}.
In the presence of parametric driving, the attenuation lengths follow a qualitatively similar dependence on the probe frequency as in equilibrium.

\bibliography{biblio}